# A Review on Large Language Models for Visual Analytics

Navya Sonal Agarwal and Sanjay Kumar Sonbhadra

*Abstract*—This paper provides a comprehensive review of the integration of Large Language Models (LLMs) with visual analytics, addressing their foundational concepts, capabilities, and wide-ranging applications. It begins by outlining the theoretical underpinnings of visual analytics and the transformative potential of LLMs, specifically focusing on their roles in natural language understanding, natural language generation, dialogue systems, and text-to-media transformations. The review further investigates how the synergy between LLMs and visual analytics enhances data interpretation, visualization techniques, and interactive exploration capabilities. Key tools and platforms including LIDA, Chat2VIS, Julius AI, and Zoho Analytics, along with specialized multimodal models such as ChartLlama and CharXIV, are critically evaluated. The paper discusses their functionalities, strengths, and limitations in supporting data exploration, visualization enhancement, automated reporting, and insight extraction. The taxonomy of LLM tasks, ranging from natural language understanding (NLU), natural language generation (NLG), to dialogue systems and text-to-media transformations, is systematically explored. This review provides a SWOT analysis of integrating Large Language Models (LLMs) with visual analytics, highlighting strengths like accessibility and flexibility, weaknesses such as computational demands and biases, opportunities in multimodal integration and user collaboration, and threats including privacy concerns and skill degradation. It emphasizes addressing ethical considerations and methodological improvements for effective integration.

## I. INTRODUCTION

LARGE Language Models (LLMs) are advanced deep learning architectures trained on vast datasets demonstrating remarkable capabilities across diverse applications, including natural language understanding, text generation, and multimodal learning [1] [2]. The rapid advancements in transformer-based architectures, particularly self-attention mechanisms, have enabled LLMs to achieve near-human proficiency in language tasks such as text generation, translation, summarization, and code synthesis [3] [4]. Beyond textual processing, recent developments in multimodal LLMs have extended their scope to integrate text, images, and audio, fostering novel applications in creative content generation, scientific research, and decision support systems [5] [6].

LLMs are deep learning driven AI models trained on extensive datasets, allowing them to recognize linguistic patterns, predict contextual meanings, and generate consistent outputs [7] [8]. These models leverage large-scale pretraining followed by fine-tuning on domain-specific datasets to enhance their performance, particularly in the fields of healthcare, finance, and law [9] [10]. Their proficiency in understanding complex queries, reasoning over large knowledge bases, and adapting to various linguistic contexts makes them invaluable in a wide range of automated systems [11] [12].

A critical advantage of LLMs is their ability to generate logical and contextually appropriate responses, facilitating advancements in conversational AI, multilingual communication, and interactive decision making [9]. Furthermore, with the integration of reinforcement learning and human feedback mechanisms, modern LLMs have demonstrated enhanced alignment with human intentions, improving accuracy, interpretability, and ethical alignment in AI deployment [13] [14]. As these models continue to evolve, their role in shaping human-AI collaboration, automated reasoning, and real-time data synthesis is expected to expand, highlighting the ongoing intersection of deep learning, cognitive computing, and visual analytics [2] [15] [16].

### A. Applications and potential of LLMs

LLMs can handle extensive quantities of data, encompassing unstructured language, and comprehend the semantic connections between words and phrases [7] [17]. These models are capable of processing visual data [18], auditory data [19], audiovisual data [20], and multimodal data [21]. They can learn the semantic links between these different types of modalities. These models have significantly improved machine's ability to comprehend and generate language that closely resembles human speech [22].

The applications of LLMs are extensive and continuously evolving. They are already causing significant impact in diverse domains, including:

- Education: AI powered tutors can provide personalized learning experiences by adapting to your student needs and answering questions in real-time [23] [24].
- Customer service: Chatbots powered by LLMs can provide more natural and helpful interactions, improving customer satisfaction [25].
- Creative industries: From writing compelling marketing campaigns to generating story ideas, LLMs can assist and inspire creative professionals [26].
- Scientific research: Analyzing vast datasets of scientific papers or generating hypotheses are just a few ways LLMs can accelerate scientific discovery [27].

With responsible development and thoughtful application, LLMs have the potential to revolutionize the way we interact with technology, learn, create, and connect. They offer a glimpse into a future where artificial intelligence not only assists us but also understands and enriches our lives in ways we can only begin to imagine. The objective of present literature review is to explore the possibilities of LLMs for visual analytics, therefore, the following subsection talks about visual analytics.

### B. Visual Analytics

In the modern data-driven realm, information is generated from multiple sources at an unprecedented scale. However, raw numerical data, figures, and statistical outputs often remain incomprehensible without appropriate analytical frameworks. Visual Analytics (VA) has emerged as a crucial interdisciplinary field that leverages human cognitive capabilities in conjunction with computational techniques to transform complex, unstructured data into meaningful insights [10]. By integrating principles from data visualization, machine learning, and human-computer interaction, VA enables users to explore, interpret, and extract actionable knowledge from large-scale datasets [28].

Visual Analytics facilitates interactive data exploration, where users can engage with large and complex datasets through intuitive visual representations. Unlike traditional static charts and graphs, VA systems incorporate real-time interaction, adaptive visualizations, and automated analytical techniques, enhancing both exploratory analysis and decision-making processes [29]. This interdisciplinary approach combines the intuitive perception of visual patterns with computational algorithms, allowing researchers and analysts to detect trends, outliers, correlations, and anomalies that may not be evident through numerical analysis alone.

It stands as a vital tool in the era of big data, enhancing our ability to process and comprehend vast amounts of information effectively. The benefits of visual analytics are manifold as follows:

- Clarity and conciseness: Visualizations condense complex data into easily digestible formats, making it accessible to a wider audience, even those without extensive statistical expertise.
- Pattern recognition: Humans excel at recognizing patterns in visual data. Charts and graphs reveal trends, outliers, and correlations that might be buried within numerical tables.
- Storytelling: Data comes alive through visuals. A captivating graph or an interactive map can tell a compelling story, engaging viewers and leaving a lasting impression.
- Informed decision-making: By uncovering hidden insights, visual analytics empowers us to make data-driven decisions, leading to better outcomes in various fields, from business and science to healthcare and policy making.

Visual analytics is more than just static charts and graphs. Present tools offer interactive features, allowing users to drill down into data, filter information, and ask questions directly through the visualizations. Advanced techniques like heatmaps, scatter plots, and network graphs provide even deeper insights into complex relationships and hidden patterns. The integration of artificial intelligence and machine learning is opening up new possibilities, such as automated data analysis, personalized visualizations, and real-time insights. The future holds exciting possibilities for visual analytics to become an even more powerful tool for unlocking the hidden stories within our data.

The primary purpose of this review paper is to explore and synthesize the current state of research at the intersection of LLMs and visual analytics. This interdisciplinary review aims to understand how the integration of LLMs enhances visual analytics capabilities, addressing both the challenges and opportunities presented by this convergence. Given the rapid advancements in both fields, the paper seeks to provide a comprehensive overview of current methodologies, applications, and future directions, offering valuable insights for researchers, practitioners, and stakeholders in the fields of data science, artificial intelligence, and visual analytics.

### C. Research Objectives

The present review aims to achieve the following research objectives:

- To identify and evaluate effective strategies for integrating Large Language Models (LLMs) into visual analytics to enhance data interpretation, interactivity, and collaboration while addressing ethical and bias challenges.
- To analyze the transformative role of LLM-powered tools in interactive data exploration and visualization by assessing their comparative advantages and limitations in real-world applications.
- To investigate the extent to which LLMs can automate insight extraction and reporting in visual analytics, focusing on natural language-based interpretation, captioning, and explainable visualization recommendations.
- To assess the improvements that LLMs bring to text-to-visualization workflows and examine the challenges in benchmarking their effectiveness for automated visual analytics.
- To conduct a SWOT analysis of LLM integration in visual data analytics and propose strategies for mitigating potential risks while maximizing benefits.

### D. Scope of the paper

Figure 1, illustrates the scope of the review including sequential organization and logical flow of the paper, demonstrating how each section contributes to the overall narrative. Here in this paper, section 2, provides background details, theoretical foundations and related work for establishing the knowledge gaps.

Section 3, explores the applications of LLMs in visual analytics, emphasizing visual user interactions and their role in enhancing analytical workflows. It covers LLM-powered insight generation, text-to-visual translation, and integrated workflows for seamless visual analytics integration. Additionally, this section discusses benchmarking methodologies for assessing LLM performance in visual analytics, presenting state-of-the-art approaches and recent advancements. Section 4, assesses the strengths, weaknesses, opportunities, and threats (SWOT) associated with application of LLMs in the domain of visual analytics. Finally section 5, concludes the paper by summarizing key findings, discussing their implications, and suggesting potential directions for future research.

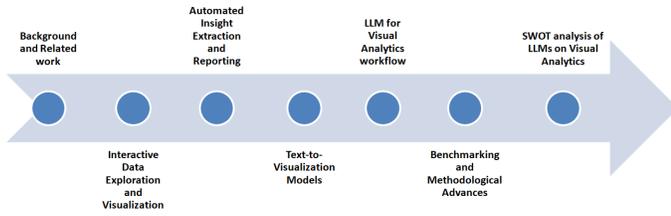

Fig. 1: Scope of the Review

## II. BACKGROUND AND RELATED WORK

This section explores the theoretical foundations of Large Language Models, presents a taxonomy for their classification, and examines their intersection with the field of Visual Analytics. It delves into their theoretical framework and explores the ongoing developments. It further highlights their evolution, methodologies, and key advancements. In addition, it also discusses about the relevance of these innovations with a strong foundation for understanding the integration of LLMs with VA for enhanced data interpretation, interactive visualization, and automated insight generation.

### A. Theoretical Foundations

*1) **Large Language Models** :* It is evident from several investigations that processing human languages through computational models were always challenging. his leads to the emergence of the Large Language Models, a subdomain of Natural Language Processing (NLP). Large Language Models are advanced artificial intelligence systems designed to process, understand, and generate human language by learning from vast datasets, significantly enhancing the capabilities of natural language processing. This enables the LLMs to execute a variety of complex tasks such as translating languages, creating content, question answering, summarization, chatbot interactions, and sentiment analysis.

The figure 3, illustrates the evolution of language models from statistical approaches to modern Large Language Models, highlighting key advancements across different eras. The timeline begins in the 1990s with N-Grams, representing the foundation of statistical language models [30]. In 2000, the transition to neural language models marked a shift towards deep learning-based methods [31]. The emergence of Word2Vec in 2013 introduced vector-based word representations [32], followed by Recurrent Neural Networks (RNNs) and Long Short-Term Memory (LSTM) models in 2014, which improved sequential data processing [33] [34]. The Attention Mechanism in 2015 further enhanced the ability of models to focus on relevant context, paving the way for modern transformer architectures [35] [3]. The introduction of Transformers in 2017, as shown in figure 2, revolutionized natural language processing (NLP), leading to the development of BERT [4] and GPT models [36] in 2018, which significantly improved contextual language understanding. This era gave rise to pretrained language models, setting the stage for the large-scale LLM revolution.

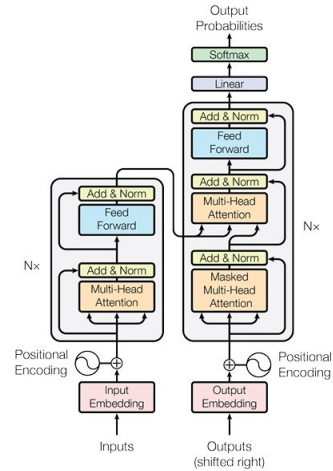

Fig. 2: Attention mechanism in transformer architecture.

The timeline further illustrates the evolution of Large Language Models (LLMs) from 2019 to 2024, showcasing major advancements in AI-driven natural language processing. It begins with Google's T5 (2019) [8] and mT5 (2020) [37], followed by OpenAI's GPT-3 (2020) [1], which significantly advanced text generation. In 2021, models like WebGPT [38], PanGu-$\alpha$ [39], and CPM-2 [40] introduced instruction tuning and web-integrated capabilities. The year 2022 saw an explosion of models, including Codex for code generation [41], Gopher [42], LaMDA [43], ChatGPT [44], and numerous multilingual models. By 2023, open-source alternatives like LLaMA [45], Vicuna [46], Koala [47], and Wizard-LM emerged alongside proprietary models like PaLM [12], BLOOM [48], and Bard [49]. The trend continued in 2024 with GPT-4 [50], Claude [51], BloombergGPT [52], and Gemini [49], emphasizing multimodal AI capabilities. The progression highlights major breakthroughs in LLMs, starting with GPT-2 in 2019, followed by GPT-3 in 2020, which demonstrated unprecedented text generation capabilities. The rapid innovation reflects the growing demand for scalable, adaptable AI models, bridging gaps between research, enterprise applications, and consumer tools. The progression continued with the release of ChatGPT in 2022, GPT-4 in 2023, and GPT-4o in 2024, showcasing increasingly sophisticated models capable of advanced reasoning, multimodal processing, and human-like interactions.

Generative AI and Large Language Models are intrinsically linked, both rooted in advanced deep learning technologies that enable content generation across various domains. LLMs, such as ChatGPT by OpenAI, Bard by Google, and Llama by Meta, are types of Generative AI models, specifically designed to generate human-like language in response to a given prompt [53]. Generative Pre-trained Transformer (GPT) series, represent a specific application of generative AI focused on text generation. As per recent developments, generative AI expands to multiple modalities such as images, audios, and videos, utilizing architectures like Generative Adversarial Networks (GANs) and Variational Autoencoders (VAEs). Here, GANs are used to improve content quality through adversarial training between a generator and a discriminator. Likewise, VAEs enable efficient latent-space encoding for various types of data generation. LLMs became more adaptable due to its transfer learning capabilities ranging from fine-tune for specific applications to improving performance without extensive task-specific data. However, the scalability of LLMs demand significant computational resources. It is also important to note that the LLM currently possesses static knowledge which is limited to their last training update. This combination of extensive training, advanced neural network architecture, and ethical concerns makes LLMs both powerful and complex tools in the field of artificial intelligence.

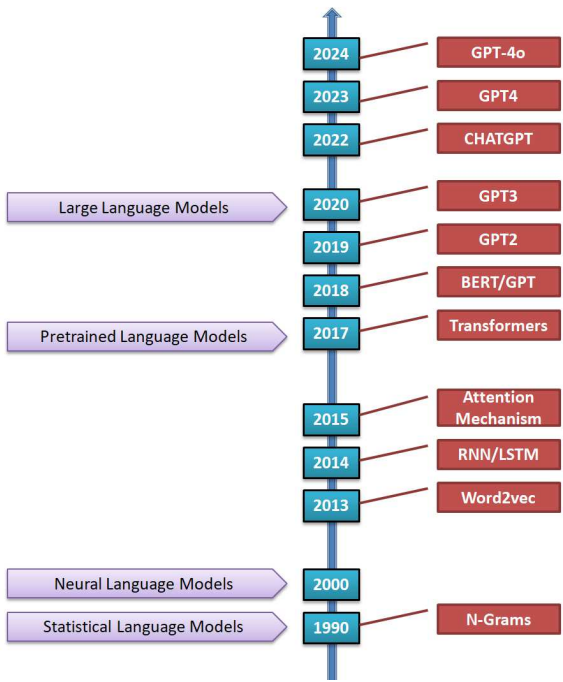

Fig. 3: Timeline of LLM Development

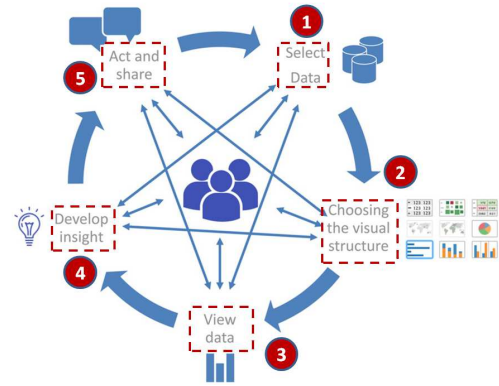

Fig. 4: A typical visual analytics approach.

*2) Concepts of Visual Analytics:* The foundation of visual analytics lies in effective representation of complex data. Visual analytics bridges the gap between data exploration and decision-making. It is an iterative process of data-driven decision-making and visualization, placing the user at the center of a continuous loop. As illustrated in figure 4, it begins with a task, followed by data collection, and the selection of appropriate visual mapping methods. The user then views the data, derives insights, and takes action by sharing or implementing findings. This interconnection of each step influences the next step, allowing for refinement and continuous improvement. This cyclic approach highlights the importance of interactive data visualization in analytics, business intelligence, and decision-making, ensuring insights evolve dynamically with new data [54]. Visual analytics also aids in hypothesis generation and testing, using visualizations to formulate and test hypotheses about underlying patterns and trends in the data.

Interactive visualization is a key aspect of visual analytics, enhancing user engagement and understanding. Dynamic querying, which allows users to interactively change query parameters and see real-time results, is a significant feature [55]. Brushing and linking, where selecting a portion of the data in one visualization affects other visualizations, enhances comparative analysis [56]. Zooming and focusing techniques enable users to delve deeper into specific data subsets, improving the granularity of analysis [57].

Handling large-scale data is a significant challenge in visual analytics. Techniques like data aggregation and sampling are essential for managing large datasets [58]. Real-time visualization is particularly relevant in stream analytics, where data is continuously updated, necessitating efficient data processing methods. Anomaly detection through visual methods helps identify outliers or unusual patterns while working with data streams [59].

Multidimensional data visualization techniques, such as parallel coordinates and scatterplot matrices, are essential for handling datasets with multiple variables [60]. For temporal data, which is particularly relevant in stream analytics, line graphs and time-series plots offer valuable insights [61]. Network and tree visualization techniques, including node-link diagrams, are crucial for understanding hierarchical and network-based data structures [62]. Additionally, geospatial data visualization, utilizing methods like choropleth maps, plays a significant role in datasets with geographical components [63].

Incorporating Human Computer Interface (HCI) principles is vital for effective visual analytics. User-centered design tailors visualizations to the target audience's needs and capabilities. Ensuring usability and accessibility is crucial for making visualizations accessible to a diverse range of users, including those with disabilities. It is also highlighted in research that managing cognitive load is also crucial in various applications. Therefore visualizations must be informative yet not overwhelming, considering human cognitive capacity [64]. Explainable AI (XAI) uses visualizations to interpret and understand the decisions made by AI models. Visualizations also assist in feature selection and model tuning, crucial for machine learning applications [65].

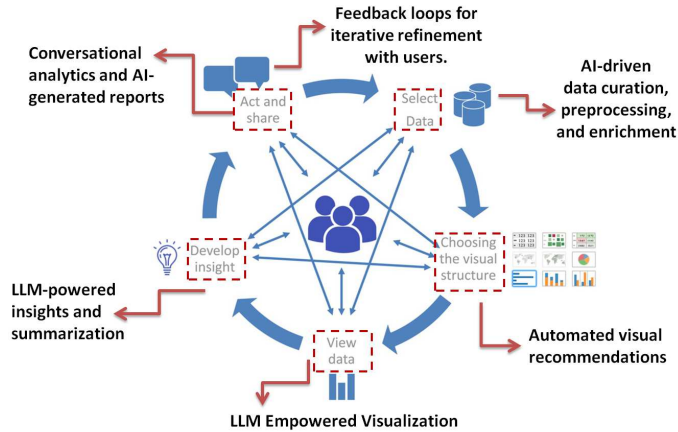

Fig. 5: Transition from traditional VA methods to recent advancements integrating LLMs.

According to recent developments, the rise of cloud-based visualization tools facilitates collaborative data exploration and analysis over distributed systems [29]. Ensuring version control and reproducibility in visual analytics processes is essential for addressing the demands of big data systems [66]. Thus, to accommodate the real time use cases of big data systems, the integration of visual analytics with AI and machine learning is emerging as a promising research area.

*3) Synergy between LLM and visual data analytics:* To meet the present demands of data-driven decision-making, the transition from traditional visual analytics methods to recent advancements integrating Large Language Models has become the need of the hour. Figure 5, illustrates the LLM-empowered visualization cycle and highlights the role of LLMs for improved visual analytics through automation, interaction, and iterative refinement. The process begins with AI-driven data curation (using tools like Selenium), preprocessing, and enrichment, ensuring data readiness.

At the beginning, users select the data and then LLMs assist in choosing the visual structure by recommending appropriate visualizations. Once the data is viewed, LLMs aid in developing insights, identifying patterns, and summarizing key findings. By generating narratives or summaries, these models can simplify the understanding of intricate patterns and insights [67]. Additionally, LLMs can provide a detailed contextual analysis of data, including interesting historical comparisons and trend analysis [68]. Through conversational analytics, users interact with LLMs for AI-generated reports, making data interpretation accessible even to non-experts [69]. Insights can be acted upon and shared, with feedback loops enabling iterative refinement to enhance decision-making. This integration makes data analysis more intuitive, efficient, and interactive, bridging the gap between raw data and actionable insights [70] [71].

With the integration of visual analytics, LLMs also support predictive analysis, offering interesting foresight into trends and facilitating proactive decision-making [72]. Through collaborative settings, it is possible to interpret and respond to queries from multiple users to facilitate multi-user interaction and knowledge sharing [73]. These features of LLM-empowered visual analytics are particularly beneficial for collaborative research and serve the needs of a wider community [74].

Likewise, the combination of LLMs and visual analytics is presented as a powerful educational tool in academic settings. It will be helpful to demonstrate complex concepts for training purposes [75]. LLMs are also useful for guided learning experiences, offering suggestions and educational content based on user interactions [76]. Additionally, LLMs help individuals with visual impairments through natural language explanations and summaries, making them proficient in data interpretation [77]. LLMs can learn from user interactions and preferences to offer personalized visualization experiences tailored to individual users' needs [78]. They can also recognize potential biases in data for better handling and mitigation [79].

There are many studies available on visual large language models for generalized and specialized applications [80]. In one such case, an application utilizes embedding of sparse causal information into video-based LLMs for enhanced learning efficiency. This research presents novel training approaches to develop the understanding of sequential visual data to facilitate the development of AI-driven video analysis. All above examples demonstrate that the synergy between LLMs and visual analytics is revolutionizing data interpretation by enhancing accessibility, improving decision-making, and enabling deeper insights across various domains.

### B. Classification of LLM Capabilities in Visual Analytics

The LLMs are widely used for visual analytics and are broadly categorized based on the nature and complexities of the applications. The figure 6, highlights the taxonomy of visual analytics tasks supported by LLMs:

- **Natural Language Understanding (NLU) Tasks:** These types of tasks help to understand and interact with the system like humans. The following are the NLU-related tasks:
    - **Text Classification:** Classifying text into predefined classes or topics [81].
    - **Sentiment Analysis:** Understanding the sentiment expressed in the text and classifying it as positive, negative, or neutral [82].
    - **Named Entity Recognition (NER):** Identifying named entities such as people, organizations, and locations in text [83].
    - **Part-of-Speech Tagging:** Extracting parts of speech (nouns, verbs, adjectives, etc.) in sentences [84].
    - **Language Inference:** Formulating relationships between sentences, such as entailment, contradiction, or neutrality [85].
    - **Question Answering:** Responding to questions based on available data or general knowledge [86].
- **Natural Language Generation (NLG) Tasks:** These types of tasks are essential to explain any context during visualization:

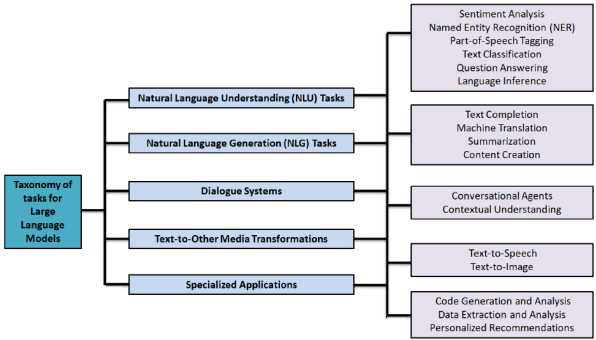

Fig. 6: Taxonomy of tasks for Large Language Models

- **Summarization:** Converting longer text into a shorter version without losing any key information [87].
- **Machine Translation:** Converting text from one language to another [88].
- **Text Completion:** Completing a given text prompt with coherent and contextually relevant content [89].
- **Content Creation:** Generating content like stories, poems, and articles [90] [91].
- **Dialogue Systems:** Having a strong dialogue system is essential to provide conversational analytics:
    - **Conversational Agents:** Establishing dialogue with users and answering user queries with appropriate visuals [92].
    - **Contextual Understanding:** Identifying the appropriate context in a conversation to generate suitable graphs and charts [93].
- **Text-to-Other Media Transformations:** Text-to-speech and text-to-image are taking visual content generation to a new dimension:
    - **Text-to-Speech:** Converting textual details into speech [94].
    - **Text-to-Image:** Generating images based on textual descriptions [95].
- **Specialized Applications:** Many professional jobs require specialized support, such as:
    - **Code Generation and Analysis:** Writing and analyzing computer code [41].
    - **Data Extraction and Analysis:** Extracting structured information from unstructured text [96].
    - **Personalized Recommendations:** Offering recommendations based on textual content analysis [97].

## III. APPLICATIONS OF LLMs IN VISUAL ANALYTICS

Large Language Models (LLMs) have significantly transformed the field of visual analytics by enabling more intuitive, interactive, and intelligent data exploration. These models enhance various aspects of data visualization, from natural language querying and automated insight generation to adaptive visual storytelling and real-time monitoring by using complex datasets. This section provides a comprehensive overview of LLMs based to visual analytics tools.

### A. Interactive Data Exploration and Visualization

Interactive data exploration and visualization play a crucial role in LLM-based visual analytics, enabling dynamic analysis and intuitive data interpretation. Below are some notable tools in this category:

*1) LIDA:* LIDA [98] is a tool for the automatic generation of grammar-agnostic visualizations and infographics using Large Language Models, significantly advancing the field of data visualization. It was launched by Microsoft and designed to simplify complex visualization tasks. LIDA consists of four core modules: Summarizer, Goal Explorer, VIZGenerator, and Infographic, which work together to streamline the data interpretation process.

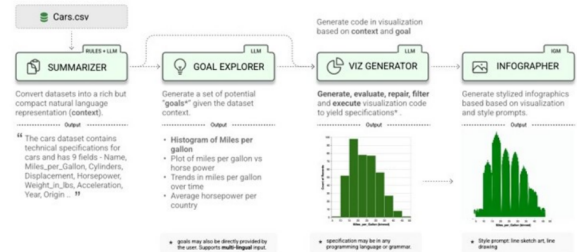

Fig. 7: Components of LIDA. Adapted from [98]

Figure 7, showcases the functionality of the four components. The summarizer module follows a two-stage process: first, generating a basic dataset summary using the pandas library, and second, enriching it with semantic information via LLM augmentation or user input. The Goal Explorer module generates data exploration objectives by treating goal generation as a multitask problem consisting of a question, a corresponding visualization, and a rationale. The VIZGenerator module comprises three submodules: a code scaffold constructor, a code generator, and a code executor. The Infographic module generates stylized graphics based on VIZ-Generator output, applying user-editable visual styles through text-conditioned image-to-image generation using diffusion models, specifically the Peacasso library API. This module offers an optional post-processing step for refining visuals by adjusting axes, removing grid lines, sharpening images, and supporting multilingual interaction and refinement. LIDA enhances accessibility, usability, and inclusivity in data visualization, making data interpretation more efficient and intuitive for a broader audience.

*2) Chat2VIS 2024:* This paper introduces a tool Chat2VIS [99] and investigates the use of LLMs for generating data visualizations from natural language queries, focusing on GPT-3, Codex, and ChatGPT. By leveraging prompt engineering, Chat2VIS effectively translates free-form user queries into visualization, making graphical representation of data more creative, more intuitive and accessible.

The tool is user-friendly, allowing users to generate visualizations effortlessly. The system provides an option to include custom datasets, where users can enter their own API key for seamless integration. Input can be provided in natural language, significantly enhancing accessibility and making it easier for users without technical expertise to interact with the system. This tool highlights the potential of LLMs in automating data visualization, reducing reliance on handcrafted rules, and streamlining the process for users across different expertise levels.

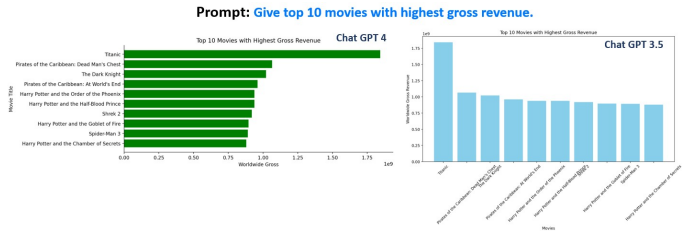

Fig. 8: Output comparison of CHAT2VIS automated visualization tool.

Figure 8, highlights the use case of Chat2VIS with the dataset movies.csv with the prompt: "Give top 10 movies with the highest gross revenue." The tool processed the request and gave results corresponding to GPT-4 and GPT-3.5. The results demonstrated that both models effectively translated the natural language query into graphs, accurately ranking the top 10 movies by revenue.

*3) Julius AI:* Julius AI [100] is an advanced data analytics platform that facilitates complex data interpretation and visualization using multiple LLMs. It offers ready-to-use workflow templates designed for specific data analysis tasks. It allows users to select and integrate different models for optimized analytical performance. Furthermore, Julius AI supports dynamic representation of data trends and patterns using animated chart options. It includes many powerful tools such as data cleaner, significance testing, time series analysis, and table extractor from PDFs and helps users to work with visualization generation while benefiting from LLM-powered automation.

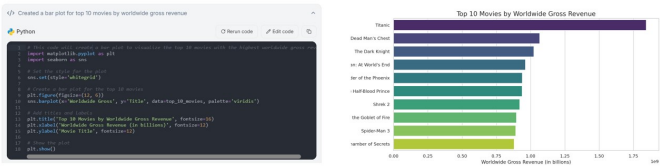

Fig. 9: On the left, the code generated by Julius AI for the given prompt is displayed, while on the right, the corresponding executed graph is shown based on the generated code

Figure 9, highlights the use of the quick visualization tool in Julius AI, where we have used the movies.csv dataset and provided the prompt "Give top 10 movies with the highest gross revenue." The tool successfully generated the corresponding visualization code along with the executed graph, effectively translating the natural language query into a structured visual representation.

*4) Zoho Analytics:* Zoho Analytics [101] is a self-service business intelligence (BI) and data analytics platform that enables users to visualize, analyze, and gain actionable insights from their data. It offers a no-code, drag-and-drop interface for creating interactive dashboards, making it accessible to both technical and non-technical users. Zoho Analytics is powered by AI-driven insights and predictive analytics, enabling businesses to detect trends, forecast outcomes, and optimize decision-making. Its collaborative reporting features allow teams to work together in real-time, enhancing data-driven strategies across departments. The platform also supports natural language querying (NLQ), allowing users to ask questions in plain English and receive instant analytical responses.

After taking the dataset as input, it automatically generated dashboard suggestions in four categories, providing a structured analysis of the dataset. Each dashboard was self-explanatory, offering a comprehensive overview of the entire dataset with visual summaries, key insights, and trends. Figure 10, gives the sample dashboard out of the four recommended dashboards, generated using the same dataset as used earlier. This Production Budget Analysis Dashboard provides insights into content rating-wise budget distribution, genre-wise allocations, top 10 titles by production budget, and creative type-based spending, helping users analyze financial trends in film production efficiently.

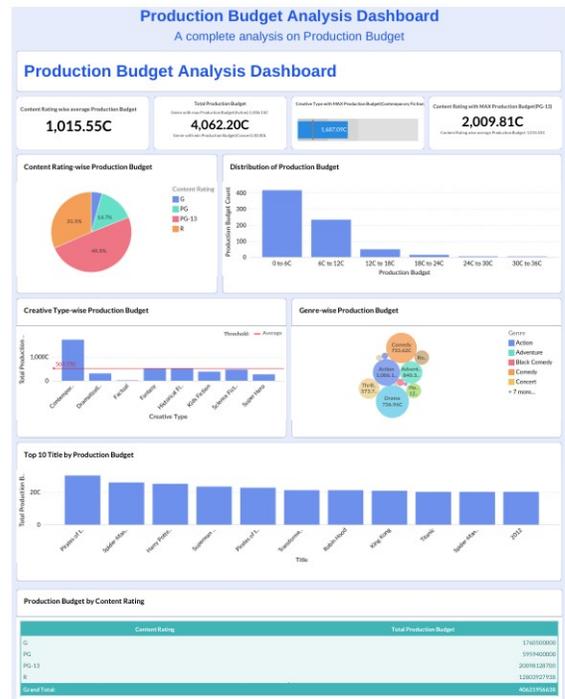

Fig. 10: Sample dashboard produced by the tool.

*5) Other tools:* **Draco 2:** is an extensible platform to model visualization design, which significantly contributes to the field of visualization design [102]. Draco 2 addresses the limitations of its predecessor by introducing a more flexible visualization specification format, comprehensive testing, documentation, and user-friendly APIs. This platform supports complex designs, including multi-layer and multi-view visualizations, and is made available as open-source, fostering collaboration and innovation in visualization research. The extensibility and adaptability of Draco 2 make it a valuable tool for researchers and practitioners seeking to integrate advanced visualization techniques into their work.

**MatplotLLM:** The development of MatplotLLM, as discussed in the EmacsConf 2023 talks, introduces a novel approach to data visualization by overlaying a natural language processing layer over the widely-used Matplotlib library [103]. This system was designed to simplify the process of creating visualizations, allowing users to leverage natural language commands within the system. The integration of LLMs with traditional visualization tools like Matplotlib represents a significant step towards making data visualization more accessible and efficient.

**Tableau AI:** Tableau AI's integration of generative AI technology, as part of Salesforce's suite of capabilities, marks a transformative shift in data analysis and insight consumption [104]. The platform offers trusted predictive and generative AI capabilities, enhancing data analysis with smart suggestions, in-product guidance, and personalized insights. This development is particularly noteworthy for its emphasis on data security, privacy, and ethical considerations in the deployment of AI technologies in business environments.

**LightVA:** It introduces a lightweight visual analytics framework [105] that integrates LLMs for agent-based task planning and execution. It aims to reduce development costs and enhance interactive data exploration by utilizing LLMs for goal-oriented task decomposition, visualization generation, and insight extraction. The system integrates a planner, executor, and controller, streamlining visual analytics workflows and improving collaboration between AI agents and human users. Table I, summarizes various features of all the tools discussed in this section.

### B. Automated Insight Extraction and Reporting

This section explores natural language-based data interpretation and captioning tools, multimodal large language models for chart understanding, and the methods used to interpret and validate AI-assisted data analyses.

*1) Natural Language-Based Data Interpretation and Captioning:* A paper based on caption generation for data visualizations explores the use of LLMs, particularly GPT-3, for generating engaging captions for data visualizations [106]. The authors argue that traditional captions often lack engagement and merely state observations. By leveraging prompt engineering techniques, the study examines how LLMs can produce more compelling captions that not only describe the visualized data but also provide journalistic insights. The paper evaluates LLM-generated captions for data visualizations through a user study, categorizing them into three tiers: Tier 1 (Template-Only, minimal detail), Tier 2 (Template+Instruction, with key talking points), and Tier 3 (Template+Instruction+QA, incorporating user modifications and clarifications). The results indicate that Tier 2 and Tier 3 captions were ranked higher for engagement and relevance due to their detailed explanations, while Tier 1 was the least preferred due to its lack of depth. Interestingly, while Tier 3 contained the most information, it sometimes introduced redundancy or unnecessary details.

Another paper examines the role of LLMs in generating natural language descriptions for data visualizations [107]. It discusses different approaches, including template-based and deep learning-based text generation methods, and evaluates their effectiveness in improving accessibility and comprehension. The study finds that LLMs can generate informative captions that enhance the interpretability of visual data, making it more accessible to users with varying levels of expertise.

A study based on chart interpretation with LLMs investigates how LLMs can aid users in interpreting complex visualizations, particularly for individuals with low data literacy [108]. It introduces a hybrid approach that integrates both textual and visual interaction to assist users in understanding complex charts. The findings highlight that while LLM-based guidance is beneficial in facilitating interpretation, it also reduces user engagement with the data, leading to overdependency on the model responses. Thus, the research explores the need for balancing AI assistance with active human interaction in data visualization tools.

*2) Multimodal Large Language Models for Chart Understanding:* ChartLlama is a multimodal LLM designed to enhance chart understanding and generation [109]. This paper proposes an instruction-tuning dataset, using GPT-4, to train models for more accurate chart interpretation and generation. The study introduces a multi-step data generation process, where different steps handle tabular data creation, chart figure generation, and instruction tuning data design separately. Experimental results demonstrate that ChartLlama outperforms existing models in tasks such as chart-to-text conversion, chart extraction, and QA-based chart interpretation. The key contribution is a comprehensive pipeline for generating diverse and high-quality chart-related data. This significantly enhances multimodal LLM's capability to process and generate charts effectively.

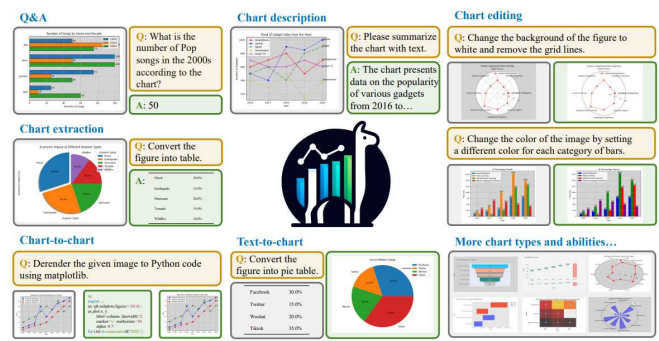

Fig. 11: Capability demonstration of ChartLlama, Adapted from [109]

Figure 11, showcases the diverse capabilities of ChartLlama in handling chart-related tasks, including Q&A, chart description, extraction, editing, text-to-chart conversion, and chart-to-chart transformations. The model efficiently interprets visual data, modifies chart attributes, and generates code for reproducing visualizations, demonstrating its multimodal proficiency in chart understanding and generation.

In another study, CharXIV introduces a comprehensive benchmark designed to assess the ability of multimodal LLMs to analyze and reason about charts [110]. It features a dataset of 2,323 real-world charts extracted from arXiv papers, ensuring a diverse, complex, and natural test set. The benchmark is structured around two types of questions: Descriptive questions, which focus on recognizing and understanding fundamental chart elements, and Reasoning questions, which require models to synthesize insights from multiple visual components. To maintain high-quality evaluation standards, all charts and questions have been carefully curated, hand-picked, and verified by human experts. The study uncovers a notable performance disparity between proprietary and open-source models, with GPT-4o achieving 47.1% accuracy, while the best-performing open-source model, InternVL Chat V1.5, reaches only 29.2%. Despite this, both models significantly underperform compared to human accuracy, which stands at 80.5%, highlighting the current limitations of multimodal LLMs in complex chart interpretation. This gap emphasizes the need for better fine-tuning and more diverse datasets to improve model reasoning abilities.

TABLE I: Comparison of Interactive Data Exploration and Visualization Tools

| Feature | LIDA | Chat2VIS | Julius AI | Zoho Analytics | Draco 2 | MatplotLLM | Tableau AI | LightVA |
|---|---|---|---|---|---|---|---|---|
| Purpose | Infographics and visualization generation | NLP-based visualization | LLM-based analytics | Self-service BI analytics | Visualization modeling | NLP-driven plots | AI-driven BI insights | AI visual analytics |
| Core Modules | Summarizer, Goal Explorer, VIZGenerator, Infographic | NLP Interface | Quick Visualization, Data Cleaner | BI, Predictive, NLQ | Constraint-based modeling | NLP layer on Matplotlib | Predictive & Generative AI | Planner, Executor, Controller |
| Customization | High (Diffusion models) | Moderate (prompt-based) | Moderate (templates) | High (interactive) | High (flexible/extensible) | Moderate (prompt-based) | High (AI-enhanced) | Moderate (AI-assisted) |
| Accessibility | High | High | High | High | Moderate | High | High | High |
| Languages/ Integration | Python, Vega-Lite, Matplotlib, Plotly, Seaborn, Bokeh | Python (GPT-based) | Multi-LLMs | BI platform (No-code) | Extensible (various frameworks) | Python, Matplotlib | Tableau ecosystem | AI-driven workflow |
| Unique Features | Diffusion-model visual refinement | Simplified NLP prompting | Quick Viz templates, Animated charts | Predictive analytics, collaboration | Advanced visualization design | NLP-commands for visualization | AI insights, ethical AI | AI-human task decomposition |
| Target Users | General, Non-experts | General, Non-experts | General, Beginners | Business analysts, Non-technical | Researchers, Experts | Beginners, General users | Business Analysts | Analysts, Researchers |

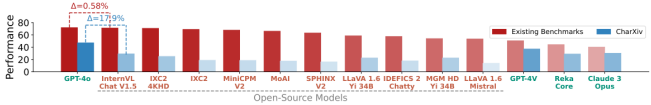

Fig. 12: Model performance comparison on reasoning questions from CharXiv v.s. questions from existing benchmarks, Adapted from [110]

Figure 12, presents a comparative analysis of model performance on reasoning questions from CharXIV versus existing benchmarks. The red and blue bars indicate that many open-source models outperform proprietary ones on 174 sample questions sourced from prior benchmarks. However, these same models struggle significantly on the 1,000 reasoning questions from the CharXIV validation split, reinforcing the higher complexity and difficulty of real-world chart reasoning compared to earlier datasets.

LLM4Vis, a ChatGPT-based framework available for recommending and explaining data visualizations [97]. The method involves feature extraction, explanation bootstrapping, and in-context learning to generate visualization recommendations with human-like justifications. The study compares LLM4Vis with traditional machine learning models and finds that it offers superior interpretability and accuracy. The research highlights the potential of LLM-driven visualization recommendations in making data exploration more accessible and insightful.

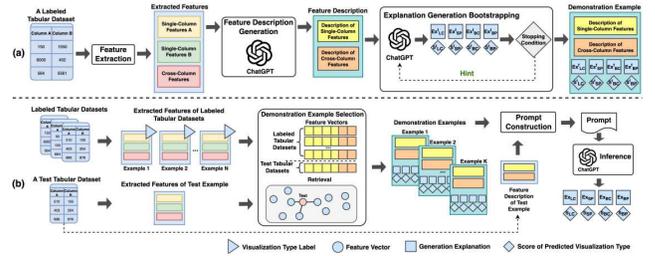

Fig. 13: A detailed illustration of LLM4Vis, Adapted from [97]

Figure 13 illustrates the LLM4Vis framework, which leverages ChatGPT for feature extraction, explanation generation, and visualization type prediction. In part (a), a labeled tabular dataset undergoes feature extraction, identifying single-column and cross-column features, which are then processed by ChatGPT to generate structured feature descriptions. Using an explanation generation bootstrapping mechanism, ChatGPT refines descriptions iteratively until a stopping condition is met, producing structured demonstration examples. In part (b), the system applies these extracted features to both labeled and test tabular datasets, generating feature vectors that facilitate the retrieval of relevant demonstration examples. These examples are then used to construct prompts containing structured descriptions of test features, which are fed into ChatGPT for inference. The final output includes visualization type labels, feature vectors, generation explanations, and scores of predicted visualization types, ensuring that LLM-driven visual recommendations are interpretable and contextually relevant.

*3) How Analysts Understand and Verify AI-Assisted Data Analyses:* This study investigates how data analysts validate AI-generated analyses, particularly in the context of AI-powered assistants that translate natural language instructions into code. The research highlights common verification behaviors, including cross-checking code, examining data transformations, and assessing AI-generated explanations. A user study reveals that analysts often shift from procedure-oriented verification to data-oriented verification when inconsistencies arise. The paper provides recommendations for improving AI-assisted analysis tools to enhance transparency and user trust.

To explore the critical aspect of validating AI-assisted data analysis Gu et al. [111] investigated the verification practices of analysts with varying backgrounds and expertise levels, focusing on how they ensure the correctness of AI-generated analyses. Through a qualitative user study involving 22 analysts, the authors uncover patterns in verification workflows and the potential misalignment between AI-assistant responses and the analyst's intent.

The study on verification processes in data analysis highlighted several limitations and proposed future research directions. It focused mainly on data transformation, omitting other key areas like data visualization, statistical modeling, and machine learning. The study also used a limited set of tasks with responses from advanced models, without controlling for the AI's assumptions or specific data operations, suggesting a need for more controlled future experiments. Additionally, participants worked with predefined analysis goals rather than their own, indicating a potential area for future research on verification processes in more personalized analysis settings with AI assistance.

### C. Text-to-Visualization Models

This section highlights various Text-to-Visualization models to showcase the capabilities of LLM supported visual analytics tools.

ChartifyText is a novel system designed to automatically convert textual data into meaningful visual representations [112]. This paper outlines a two-stage process involving tabular data inference and expressive chart generation. The proposed model leverages LLMs like GPT-4 to extract and structure data, while advanced visualization techniques ensure that charts effectively convey the underlying information.

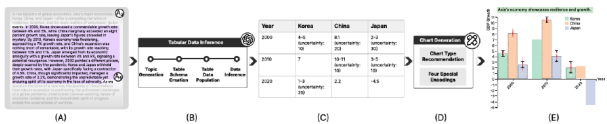

Fig. 14: A step by step breakdown of graph generation using ChartifyText, Adapted from [112]

Figure 14, illustrates the process of transforming unstructured text into a structured data visualization. The pipeline involves (A) extracting key insights from textual data, (B) inferring tabular data structure, (C) generating a structured table with uncertainty annotations, (D) recommending an appropriate chart type, and (E) visualizing the final output with special encodings for better interpretability.

Another study introduces Text2Chart, a multi-stage pipeline that automatically generates visualizations from natural language text [113]. The approach relies on a combination of neural network architectures, including LSTMs and transformers, the model identifies key entities for the x- and y-axes, maps relationships between them, and selects the most appropriate chart type (bar, line, or pie chart). Th paper highlights challenges related to dataset availability and potential improvements through larger training corpora.

ChartGPT is a tool that leverages LLMs to transform natural language queries into meaningful data visualizations [114]. It integrates deep learning models with chart generation mechanisms to automate the process of visualization creation. The study explores various challenges, such as ambiguity in text input and selecting the most appropriate visualization types, demonstrating how LLMs can refine data-driven decision-making.

### D. LLM for Visual Analytics workflow

This section covers the LLM based tools designed for visual analytics workflow management. The iGAiVA framework integrates Generative AI and Visual Analytics to enhance machine learning workflows, particularly in text classification tasks [115]. This paper highlights how visual analytics can help to identify data deficiencies and guide the generation of synthetic data using large language models. By utilizing the interactive visual tools, iGAiVA enables ML developers to pinpoint gaps in training datasets and generate targeted synthetic samples, improving overall model accuracy. The proposed approach maps ML tasks into four visual analytics views, streamlining model development and refinement. The study demonstrates that integrating generative AI with visual analytics can effectively address challenges related to data scarcity and skewed distributions and contribute to real-world machine learning applications.

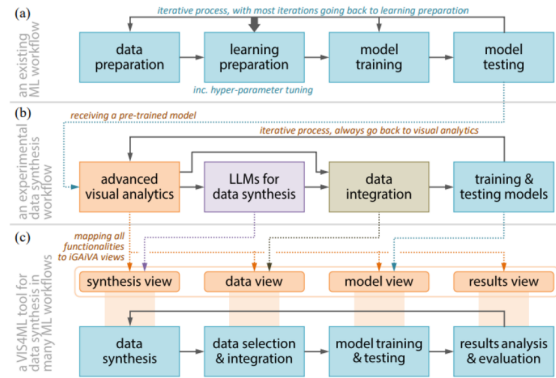

Fig. 15: VIS4ML: LLM for Visual Analytics workflow, Adapted from [115].

Figure 15, presents a comparison between traditional machine learning workflow and an experimental workflow: integrating LLMs for data synthesis, and a VIS4ML tool mapping functionalities to various ML workflow stages. The traditional ML workflow follows a linear process of data preparation, learning, model training, and testing, with iterative refinements. This experimental workflow enhances this process by incorporating advanced visual analytics and LLM-powered data synthesis, enabling better data integration and iterative improvements. The VIS4ML tool further refines this by structuring the workflow into synthesis, data, model, and results views, facilitating efficient data synthesis, selection, model training, and evaluation for improved ML outcomes.

In another study, LLM-Enhanced Visual Analytics (LEVA) introduces a framework that utilizes LLMs to enhance various stages of visual analytics workflows, including on-boarding, exploration, and summarization [116]. The system assists its users by interpreting visualization designs, suggesting data insights based on system interactions, and automatically generating structured reports. By embedding LLM-powered recommendations and mixed-initiative exploration, LEVA aims to bridge the gap between users and complex VA tools, making data interpretation more efficient. The paper presents usage scenarios and a user study to validate the framework's effectiveness, demonstrating how LLMs can facilitate intelligent interactions and improve user experience in visual analytics.

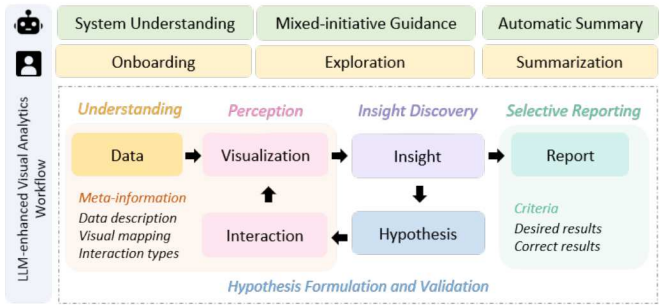

Fig. 16: The LLM-enhanced visual analytics workflow using LEVA framework, Adapted from [116].

Figure 16 illustrates the LLM-enhanced visual analytics workflow, integrating system understanding, mixed-initiative guidance, and automatic summarization. The process begins with on-boarding, where users interact with data through visualization and interaction mechanisms. This leads to insight discovery, where data is analyzed, hypotheses are formulated, and meaningful insights are derived. Finally, in the selective reporting phase, relevant insights are compiled into a structured report based on specific criteria. The framework ensures a seamless transition from data exploration to insight generation, improving the interpretability and usability of visual analytics systems.

### E. Benchmarking and Methodological Advances

Recent studies have further expanded the capabilities of LLMs in visual analytics. Benchmark datasets include CharXIV [110], ChartQA [117], and FigureQA [118], each designed to evaluate different aspects of chart comprehension and reasoning. CharXIV consists of 2,323 real-world charts from arXiv papers, incorporating both descriptive and reasoning-based questions, ensuring a diverse and challenging evaluation. ChartQA is another structured dataset that focuses on question answering over charts, requiring models to extract numerical insights and understand relationships within visual data. FigureQA assesses visual reasoning abilities by presenting automatically generated figures with binary question-answer pairs, testing the LLM's ability to interpret color, shape, and trend-based information in visual analytics.

Luo et al. [119] introduced an innovative framework nvBench, using synthesizing natural language to visualization (NL2VIS) benchmarks from NL2SQL benchmarks that represents a significant methodological advancement in the field of data visualization. This framework was designed to synthesize benchmarks for natural language to visualization tasks, leveraging the relationship between SQL queries and visualization queries. This framework not only facilitates the advancement of data visualization through deep learning models but also underscores the evolving synergy between natural language processing and visual data representation. This study's approach utilizes a unified intermediate representation, especially for its efficiency in benchmark development, potentially reducing the time and effort typically required in this process. The implications of this study are far-reaching, offering a streamlined pathway for future research and development in the field of NL2VIS.

Li et al. [120] provide an insightful examination of the integration of artificial intelligence in data storytelling tools. This study is pivotal in understanding the collaborative dynamics between humans and AI in the storytelling process. The authors meticulously analyze existing tools through a comprehensive framework that considers both the stages of storytelling (analysis, planning, implementation, and communication) and the roles of humans and AI (creators, assistants, optimizers, reviewers) in each stage. The study identifies common collaboration patterns, summarizes lessons learned, and outlines research opportunities for human-AI collaboration in data storytelling.

Hong and Crisan [121] explored the application of LLMs in enhancing data visualization through conversational interfaces in their study. This research, focusing on chatbots in visual analysis, led to the development of AI Threads, a multi-threaded analytic chatbot designed to manage conversational context and output efficiency. Leonardo and Paulovich [122] proposed ChatKG for analyzing temporal data through knowledge graphs, utilizing chat AI like ChatGPT. This method allows for the exploration of patterns in datasets like world life expectancy, correlating them with relevant information extracted via ChatGPT. Trebuňa et al. [123] presented VisuaLLM, a Python library that facilitates interactive visualization of natural language generation tasks using pretrained language models. This tool, integrating with the HuggingFace API, offers a user-friendly web interface for visualizing data, predicting next tokens, and controlling decoding parameters, thereby aiding in the understanding and evaluation of language model behaviors.

Another paper [124] examines the interpretative abilities of multimodal LLMs in visual tasks. The study explores how these models understand, generate, and reason about visual data, highlighting areas where human-like comprehension remains a challenge. Likewise, another paper investigates how multimodal models process images and videos, with a focus on their ability to perform causal reasoning and make intuitive inferences about visual content [125].

PromptAid [126] explores the complexities of utilizing LLMs for Natural Language Processing tasks. It is a visual analytics system designed to interactively create, refine, and test prompts through exploration, perturbation, testing, and iteration. The study emphasizes the challenges faced by non-expert users in crafting effective prompts to guide LLMs. The authors argue for the necessity of user-friendly interfaces that can assist users in navigating the intricacies of prompt engineering as shown in Fig. 17. This paper is pivotal in understanding the gap between LLM capabilities and user proficiency, advocating for tools that simplify the interaction with these complex models.

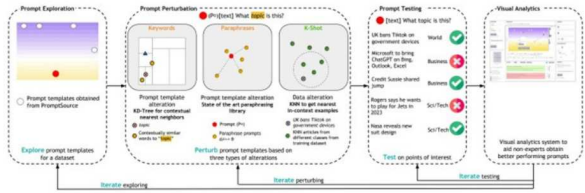

Fig. 17: Working of PromptAid.

From the above section, it is evident that advancements in benchmarking and methodological innovations continue to push the boundaries of LLM applications in visual analytics. It is also clear that as LLMs evolve, their integration with visual analytics will also achieve better intelligence and adaptiveness.

## IV. SWOT ANALYSIS OF LLMs FOR VISUAL DATA ANALYTICS

The assimilation of LLMs into Visual Analytics characterizes a transformative progression and offers potential benefits. However, it also raise notable challenges. Integration of LLMs not only significantly enhance user interactions but also facilitate scalable data management. It also improves ease of use and flexibility for users across diverse expertise levels. Nevertheless, the integration of LLMs introduces certain limitations and risks, such as accuracy concerns, computational complexity, ethical considerations, and data privacy issues. The SWOT analysis here, thoroughly examine these internal strengths and weaknesses, as well as external opportunities and threats to provide a complete viewpoint on adopting LLMs for visual analytics.

### A. Strengths

*a) Better accessibility:* Using LLMs for natural language interactions, visual analytics systems have become significantly more accessible, particularly to people who lack data science or visualization coding expertise. They can simplify this process by converting plain-text queries directly into visual analytics tasks. Overall, reducing the need for users to learn specialized tools and programming languages.

*b) Enhanced Flexibility:* LLMs offer a variety of capabilities by handling various user queries and analytical tasks in natural language. This flexibility has surpassed traditional rule-based natural language interfaces, as they usually provide only limited and narrowly defined command sets.

*c) Scalability:* Recent advancements have demonstrated that LLMs (such as GPT-4) can effectively automate data preprocessing tasks. Schema matching, data cleaning, and error detection have significantly streamlined the initial stages of data preparation and enabling scalable data management.

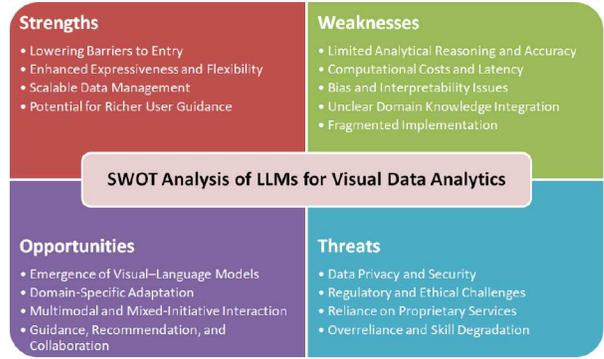

Fig. 18: SWOT Analysis

*d) Interactive User Interface:* LLMs have enriched user interactions by incorporating real-time explanations, clarifications, or suggestions. It collectively promotes an interactive, mixed-initiative approach involving users with maximum interest. In addition, the LLM models are helpful to enhance real-time analytics by generating instant summaries, identifying errors, and automated decision-making processes.

### B. Weaknesses

*a) Limited Analytical Reasoning and Accuracy:* Current LLMs occasionally can produce plausible yet incorrect facts (commonly termed as hallucinations), which undermines the reliability of analytical insights derived from these models. In addition to this, LLMs depict poor results while dealing with with complex or ambiguous user queries. The commonly reported weakness are failing to correctly interpret, thus potentially compromising analytical integrity.

*b) Computational Costs and Latency:* Large-scale LLMs involve extensive computational resources that cause high latency and significant infrastructure costs. This demands substantial model optimization by deploying smaller, specialized, and fine-tuned models tailored for specific visualization tasks. Efficient optimization techniques reduce resource demands without significantly compromising the performance.

*c) Bias and Interpretability Issues:* LLMs frequently suffer with biased results due to their imbalanced training data and cause misleading or incorrect interpretations. Therefore, it is necessary to address these biases by using suitable fairness-aware training methods. Applying appropriate post-processing techniques can also help to enhance interpretability and reduce associated risks.

*d) Black-Box Nature:* LLMs characteristically function as black-box models and lacks transparency regarding the derived insights or recommendations. This opacity seriously conflicts with the basic principle of visual analytics, which emphasizes transparent and demonstrable analysis. Therefore, introducing Explainable AI (XAI) techniques becomes mandatory, particularly in accountability-sensitive domains.

*e) Unclear Domain Knowledge Integration:* While LLMs handles extensive domain specific data, their knowledge may not be adequately specific. Integrating specific domain expertise often involves additional fine-tuning and careful designing of prompts to ensure relevancy and accuracy.

*f) Fragmented Implementation:* Presently, LLM implementations are present in the form of multiple, disconnected modules—such as separate processes for generating Python code for visualizations or distinct modules for data cleaning. These fragmented implementations seek numerous challenges for seamless integration. To resolve this issue, it is essential to develop interactive feedback loops which can unify different components and allow users to iteratively refine outputs, thus enhancing overall accuracy.

### C. Opportunities

*a) Emergence of Visual–Language Models:* The development of specialized visual-language models capable of directly interpreting or generating visual encoding without intermediary code presents significant opportunities. Additionally, such models could analyze and reverse-engineer existing visualizations from published sources, extracting deeper insights from large-scale visual data repositories.

*b) Domain-Specific Adaptation:* Fine-tuning and domain-specific enhancement of LLMs offer opportunities for more tailored, accurate, and efficient analytics across diverse fields such as healthcare, finance, and engineering. These adaptations enable more targeted predictive modeling through automated preprocessing, feature engineering, contextual analysis, and the generation of actionable, human-readable reports.

*c) Multimodal and Mixed-Initiative Interaction:* Integrating natural language interactions with traditional analytic methods—such as direct manipulation, sketching, speech, or gestures—can significantly enhance user experiences. Such mixed-initiative interfaces allow LLMs to maintain conversational context, proactively surface insights, and support dynamic real-time analytics and streaming data scenarios.

*d) Guidance, Recommendation, and Collaboration:* LLMs have the potential to actively guide analytics processes by recommending next steps, highlighting anomalies, or refining visual outputs based on real-time user interactions. Enhanced adaptive guidance further facilitates interactive visualization refinement, deepening user engagement and improving analytic outcomes.

### D. Threats

*a) Data privacy and security:* Utilizing LLMs for various analytics tasks often requires uploading sensitive or proprietary data to external cloud services. It leads potential conflicts with privacy laws and increase IP risks. This further demands to incorporate differential privacy, federated learning, and blockchain technology to bolster data protection.

*b) Regulatory and ethical challenges:* Risks of misinformation or imprecise analysis generated by LLMs elevate severe ethical and regulatory concerns, especially in areas of healthcare and finance. Furthermore, it is essential to conduct a thorough verification of the LLM generated results before its acceptances. It would be essential to address these biases early in model training through detailed bias detection and mitigation frameworks.

*c) Over-reliance and skill degradation:* Excessive dependence on automated LLM-driven systems may critically affect the role and expertise of analysts. Sometimes blind trust in system outputs without adequate understanding or human oversight could result in unnoticed anomalies and errors requiring deeper expertise. Over reliance on proprietary and closed-source LLM platforms could result reduced customization opportunities. This sometimes also causes potential disruptions due to discontinuation of vendor's services.

## V. CONCLUSION

The integration of LLMs with visual analytics is rapidly evolving, unlocking new possibilities for data exploration, interpretation, and interactivity. This review identifies key trends, including the rise of multimodal LLMs that seamlessly integrate textual and visual data for deeper insights and more comprehensive analysis. The ability of LLMs to personalize data visualization based on user interaction further enhances accessibility and engagement, making complex analytics more intuitive for a broader audience. Advancements in visual analytics tools and models have significantly improved visual analytics tasks across varying expertise levels. Multimodal approaches now support tasks such as text-to-chart generation, automated chart interpretation, and dynamic customization, enabling more efficient analytical workflows. Additionally, AI-assisted insights and enhanced human-AI collaboration ensure adaptability and effectiveness in diverse analytical applications. The SWOT analysis highlights LLMs' strengths in scalability and interactivity, alongside challenges in accuracy, computational costs, and interpretability, necessitating optimization and explainable AI solutions. It presents the emerging opportunities in visual-language models and multimodal AI models to refine analytics. Lastly it also addresses the concerns related to privacy risks, ethical standards, and over dependence on AI based solutions. Thus, as research progresses, LLMs are set to shape the next generation of intuitive, interactive, and transparent analytics tools, fostering more efficient and insightful data-driven decision-making across domains.

## REFERENCES

[1] T. Brown, B. Mann, N. Ryder, M. Subbiah, J. D. Kaplan, P. Dhariwal, A. Neelakantan, P. Shyam, G. Sastry, A. Askell *et al.*, "Language models are few-shot learners," *Advances in neural information processing systems*, vol. 33, pp. 1877–1901, 2020.

[2] R. Bommasani, D. A. Hudson, E. Adeli, R. Altman, S. Arora, S. von Arx, M. S. Bernstein, J. Bohg, A. Bosselut, E. Brunskill *et al.*, "On the opportunities and risks of foundation models," *arXiv preprint arXiv:2108.07258*, 2021.

[3] A. Vaswani, N. Shazeer, N. Parmar, J. Uszkoreit, L. Jones, A. N. Gomez, Ł. Kaiser, and I. Polosukhin, "Attention is all you need," *Advances in neural information processing systems*, vol. 30, 2017.

[4] J. Devlin, M.-W. Chang, K. Lee, and K. Toutanova, "BERT: Pre-training of Deep Bidirectional Transformers for Language Understanding," in *Proceedings of the 2019 Conference of the North American Chapter of the Association for Computational Linguistics: Human Language Technologies, Volume 1*, 2019, pp. 4171–4186.

[5] J.-B. Alayrac, J. Donahue, P. Luc, A. Miech, I. Barr, Y. Hasson, K. Lenc, and et al., "Flamingo: a Visual Language Model for Few-Shot Learning," in *Advances in Neural Information Processing Systems*, vol. 35, 2022, pp. 23 716–23 736.

[6] H. Liu, C. Li, Q. Wu, and Y. J. Lee, "Visual Instruction Tuning," in *Advances in Neural Information Processing Systems*, vol. 36, 2023, pp. 34 892–34 916.

[7] A. Radford, J. Wu, R. Child, D. Luan, D. Amodei, I. Sutskever *et al.*, "Language models are unsupervised multitask learners," *OpenAI blog*, vol. 1, no. 8, p. 9, 2019.

[8] C. Raffel, N. Shazeer, A. Roberts, K. Lee, S. Narang, M. Matena, Y. Zhou, W. Li, and P. J. Liu, "Exploring the limits of transfer learning with a unified text-to-text transformer," *Journal of Machine Learning Research*, vol. 21, pp. 1–67, 2020.

[9] A. Savikhin, R. Maciejewski, and D. S. Ebert, "Applied visual analytics for economic decision-making," in *2008 IEEE Symposium on Visual Analytics Science and Technology*, 2008, pp. 107–114.

[10] L. Zhang, A. Stoffel, M. Behrisch, S. Mittelstadt, T. Schreck, R. Pompl, S. Weber, H. Last, and D. Keim, "Visual analytics for the big data era — a comparative review of state-of-the-art commercial systems," in *2012 IEEE Conference on Visual Analytics Science and Technology (VAST)*, 2012, pp. 173–182.

[11] OpenAI, "Gpt-4 technical report," *arXiv preprint arXiv:2303.08774*, 2023.

[12] A. Chowdhery, S. Narang, J. Devlin, M. Bosma, G. Mishra, A. Roberts, P. Barham, et al., "PaLM: Scaling Language Modeling with Pathways," *Journal of Machine Learning Research*, vol. 24, no. 240, pp. 1–113, 2023.

[13] L. Ouyang, J. Wu, X. Jiang, D. Almeida, C. Wainwright, P. Mishkin, C. Zhang, S. Agarwal, K. Slama, A. Ray *et al.*, "Training language models to follow instructions with human feedback," *Advances in neural information processing systems*, vol. 35, pp. 27 730–27 744, 2022.

[14] L. Weidinger, C. Teehan, T. Gksun *et al.*, "Ethical and social risks of harm from language models," *arXiv preprint arXiv:2112.04359*, 2021. [Online]. Available: https://arxiv.org/abs/2112.04359

[15] A. Dosovitskiy, L. Beyer, A. Kolesnikov *et al.*, "An image is worth 16x16 words: Transformers for image recognition at scale," *International Conference on Learning Representations (ICLR)*, 2021. [Online]. Available: https://arxiv.org/abs/2010.11929

[16] H. Strobelt, S. Gehrmann, M. Behrisch *et al.*, "Seq2seq-vis: A visual debugging tool for sequence-to-sequence models," *IEEE Transactions on Visualization and Computer Graphics*, vol. 25, no. 1, pp. 353–363, 2018. [Online]. Available: https://arxiv.org/abs/1804.09299

[17] Y. Zhu, J. R. A. Moniz, S. Bhargava, J. Lu, D. Piraviperumal, S. Li, Y. Zhang, H. Yu, and B.-H. Tseng, "Can large language models understand context?" 2024. [Online]. Available: https://arxiv.org/abs/2402.00858

[18] M. Awais, M. Naseer, S. Khan, R. M. Anwer, H. Cholakkal, M. Shah, M.-H. Yang, and F. S. Khan, "Foundational models defining a new era in vision: A survey and outlook," *arXiv preprint arXiv:2307.13721*, 2023.

[19] H. Zhang, X. Li, and L. Bing, "Video-LLAMA: An instruction-tuned audio-visual language model for video understanding," *arXiv preprint arXiv:2306.02858*, 2023.

[20] A. Rouditchenko, A. Boggust, D. Harwath, B. Chen, D. Joshi, S. Thomas, K. Audhkhasi, H. Kuehne, R. Panda, R. Feris *et al.*, "Avlnet: Learning audio-visual language representations from instructional videos," *arXiv preprint arXiv:2006.09199*, 2020.

[21] Y. Zhao, Z. Lin, D. Zhou, Z. Huang, J. Feng, and B. Kang, "Bubogpt: Enabling visual grounding in multi-modal llms," *arXiv preprint arXiv:2307.08581*, 2023.

[22] J. Huang and K. C.-C. Chang, "Towards reasoning in large language models: A survey," *arXiv preprint arXiv:2212.10403*, 2022.

[23] S. Moore, R. Tong, A. Singh, Z. Liu, X. Hu, Y. Lu, J. Liang, C. Cao, H. Khosravi, P. Denny *et al.*, "Empowering education with llms-the next-gen interface and content generation," in *International Conference on Artificial Intelligence in Education*. Springer, 2023, pp. 32–37.

[24] B. Alhafni, S. Vajjala, S. Bannò, K. K. Maurya, and E. Kochmar, "Llms in education: Novel perspectives, challenges, and opportunities," *arXiv preprint arXiv:2409.11917*, 2024.

[25] K. Pandya and M. Holia, "Automating customer service using langchain: Building custom open-source gpt chatbot for organizations," *arXiv preprint arXiv:2310.05421*, 2023.

[26] H. Sajjad, "Generative ai: Transforming creative industries with machine learning," *Research Corridor Journal of Engineering Science*, vol. 1, no. 2, pp. 119–134, 2024.

[27] Y. Zhou, H. Liu, T. Srivastava, H. Mei, and C. Tan, "Hypothesis generation with large language models," *arXiv preprint arXiv:2404.04326*, 2024. [Online]. Available: https://arxiv.org/abs/2404.04326

[28] J. J. Thomas and K. A. Cook, "A visual analytics agenda," *IEEE computer graphics and applications*, vol. 26, no. 1, pp. 10–13, 2006.

[29] J. Heer, J. Mackinlay, C. Stolte, and M. Agrawala, "Graphical histories for visualization: Supporting analysis, communication, and evaluation," *IEEE transactions on visualization and computer graphics*, vol. 14, no. 6, pp. 1189–1196, 2008.

[30] C. E. Shannon, "A mathematical theory of communication," *The Bell System Technical Journal*, vol. 27, no. 3, pp. 379–423, 1948.

[31] Y. Bengio, R. Ducharme, P. Vincent, and C. Jauvin, "A neural probabilistic language model," *Journal of Machine Learning Research*, vol. 3, pp. 1137–1155, 2003.

[32] T. Mikolov, I. Sutskever, K. Chen, G. S. Corrado, and J. Dean, "Efficient estimation of word representations in vector space," in *Proceedings of ICLR*, 2013.

[33] S. Hochreiter and J. Schmidhuber, "Long short-term memory," *Neural Computation*, vol. 9, no. 8, pp. 1735–1780, 1997.

[34] K. Cho, B. Van Merriënboer, C. Gulcehre, D. Bahdanau, F. Bougares, H. Schwenk, and Y. Bengio, "Learning phrase representations using rnn encoder-decoder for statistical machine translation," in *Proceedings of EMNLP*, 2014.

[35] D. Bahdanau, K. Cho, and Y. Bengio, "Neural machine translation by jointly learning to align and translate," *Proceedings of ICLR*, 2015.

[36] A. Radford, K. Narasimhan, T. Salimans, I. Sutskever *et al.*, "Improving language understanding by generative pre-training," 2018.

[37] L. Xue, N. Constant, A. Roberts, M. Kale, R. Al-Rfou, A. Siddhant, A. Barua, and C. Raffel, "mt5: A massively multilingual pre-trained text-to-text transformer," *arXiv preprint arXiv:2010.11934*, 2020.

[38] R. Nakano, J. Hilton, S. Balaji, J. Wu, L. Ouyang, C. Kim, C. Hesse, S. Jain, V. Kosaraju, W. Saunders *et al.*, "Webgpt: Browser-assisted question-answering with human feedback," *arXiv preprint arXiv:2112.09332*, 2021.

[39] W. Zeng, X. Ren, T. Su, H. Wang, Y. Liao, Z. Wang, X. Jiang, Z. Yang, K. Wang, X. Zhang *et al.*, "Pangu-$\alpha$: Large-scale autoregressive pretrained chinese language models with auto-parallel computation," *arXiv preprint arXiv:2104.12369*, 2021.

[40] Z. Zhang, Y. Gu, X. Han, S. Chen, C. Xiao, Z. Sun, Y. Yao, F. Qi, J. Guan, P. Ke *et al.*, "Cpm-2: Large-scale cost-effective pre-trained language models," *AI Open*, vol. 2, pp. 216–224, 2021.

[41] M. Chen, J. Tworek, H. Jun, Q. Yuan, H. P. D. O. Pinto, J. Kaplan, H. Edwards, Y. Burda, N. Joseph, G. Brockman *et al.*, "Evaluating large language models trained on code," *arXiv preprint arXiv:2107.03374*, 2021.

[42] J. W. Rae, S. Borgeaud, T. Cai, K. Millican, J. Hoffmann, F. Song, J. Aslanides, S. Henderson, R. Ring, S. Young *et al.*, "Scaling language models: Methods, analysis & insights from training gopher," *arXiv preprint arXiv:2112.11446*, 2021.

[43] R. Thoppilan, D. D. Freitas, J. Hall, N. Shazeer, A. Kulshreshtha, A. Cheng, A. Jin, T. Bos, Q. Baker, Y. Du *et al.*, "Lamda: Language models for dialog applications," *arXiv preprint arXiv:2201.08239*, 2022.

[44] K. I. Roumeliotis and N. D. Tselikas, "Chatgpt and open-ai models: A preliminary review," *Future Internet*, vol. 15, no. 6, p. 192, 2023.

[45] H. Touvron, T. Lavril, G. Izacard, X. Martinet, M.-A. Lachaux, T. Lacroix, B. Rozière, N. Goyal, E. Hambro, F. Azhar *et al.*, "Llama: Open and efficient foundation language models," *arXiv preprint arXiv:2302.13971*, 2023.

[46] B. Peng, C. Li, P. He, M. Galley, and J. Gao, "Instruction tuning with gpt-4," *arXiv preprint arXiv:2304.03277*, 2023.

[47] R. Tan, X. Sun, P. Hu, J.-h. Wang, H. Deilamsalehy, B. A. Plummer, B. Russell, and K. Saenko, "Koala: Key frame-conditioned long video-llm," in *Proceedings of the IEEE/CVF Conference on Computer Vision and Pattern Recognition*, 2024, pp. 13 581–13 591.

[48] T. Le Scao, A. Fan, C. Akiki, E. Pavlick, S. Ilić, D. Hesslow, R. Castagné, A. S. Luccioni, F. Yvon, M. Gallé *et al.*, "Bloom: A 176b-parameter open-access multilingual language model," 2023.

[49] G. Team, R. Anil, S. Borgeaud, Y. Wu, J. Alayrac, J. Yu, R. Soricut, J. Schalkwyk, A. Dai, A. Hauth *et al.*, "Gemini: A family of highly capable multimodal models," 2024," *arXiv preprint arXiv:2312.11805*, 2024.

[50] J. Achiam, S. Adler, S. Agarwal, L. Ahmad, I. Akkaya, F. L. Aleman, D. Almeida, J. Altenschmidt, S. Altman, S. Anadkat *et al.*, "Gpt-4 technical report," *arXiv preprint arXiv:2303.08774*, 2023.

[51] Anthropic AI, "Claude AI: Conversational AI by Anthropic," https://www.anthropic.com/claude, accessed: 2025-03-15.

[52] S. Wu, O. Irsoy, S. Lu, V. Dabravolski, M. Dredze, S. Gehrmann, P. Kambadur, D. Rosenberg, and G. Mann, "Bloomberggpt: A large language model for finance," *arXiv preprint arXiv:2303.17564*, 2023.

[53] A. Jo, "The promise and peril of generative ai," *Nature*, vol. 614, no. 1, pp. 214–216, 2023.

[54] T. Ruppert, "Visual analytics to support evidence-based decision making," Ph.D. dissertation, TU Darmstadt (TUPrints), 2018. [Online]. Available: https://tuprints.ulb.tu-darmstadt.de/

[55] B. Shneiderman, "Dynamic queries for visual information seeking," *IEEE software*, vol. 11, no. 6, pp. 70–77, 1994.

[56] A. Buja, J. A. McDonald, J. Michalak, and W. Stuetzle, "Interactive data visualization using focusing and linking," in *Proceedings of the 2nd conference on Visualization'91*, 1991, pp. 156–163.

[57] J. S. Yi, Y. ah Kang, J. Stasko, and J. A. Jacko, "Toward a deeper understanding of the role of interaction in information visualization," *IEEE transactions on visualization and computer graphics*, vol. 13, no. 6, pp. 1224–1231, 2007.

[58] Y. Wang, "Deck.gl: Large-scale web-based visual analytics made easy," *arXiv preprint arXiv:1910.08865*, 2019. [Online]. Available: https://arxiv.org/abs/1910.08865

[59] F. Dinmohammadi and D. Wilson, "Understanding the end-users and technical requirements for real-time streaming data analytics and visualisation," in *2021 26th International Conference on Automation and Computing (ICAC)*. IEEE, 2021, pp. 1–6.

[60] D. Keim, J. Kohlhammer, G. Ellis, and F. Mansmann, *Mastering the information age solving problems with visual analytics*. Eurographics Association, 2010.

[61] W. Aigner, S. Miksch, H. Schumann, and C. Tominski, *Visualization of time-oriented data*. Springer, 2011, vol. 4.

[62] I. Herman, G. Melançon, and M. S. Marshall, "Graph visualization and navigation in information visualization: A survey," *IEEE Transactions on visualization and computer graphics*, vol. 6, no. 1, pp. 24–43, 2000.

[63] R. S. Bivand, "Exploratory spatial data analysis," in *Handbook of applied spatial analysis: Software tools, methods and applications*. Springer, 2009, pp. 219–254.

[64] S. Kalyuga, "Managing cognitive load in dynamic visual representations," in *Managing Cognitive Load in Adaptive Multimedia Learning*. IGI Global, 2009, pp. 171–197.

[65] G. Alicioglu and B. Sun, "A survey of visual analytics for explainable artificial intelligence methods," *Computers & Graphics*, vol. 102, pp. 502–520, 2022.

[66] E. D. Ragan, A. Endert, J. Sanyal, and J. Chen, "Characterizing provenance in visualization and data analysis: an organizational framework of provenance types and purposes," *IEEE transactions on visualization and computer graphics*, vol. 22, no. 1, pp. 31–40, 2015.

[67] E. Hoque and M. S. Islam, "Natural language generation for visualizations: State of the art, challenges and future directions," *Computer Graphics Forum*, vol. 44, no. 1, p. e15266, 2025.

[68] D. Sacha, A. Stoffel, F. Stoffel, B. C. Kwon, G. Ellis, and D. A. Keim, "Knowledge generation model for visual analytics," *IEEE Transactions on Visualization and Computer Graphics*, vol. 20, no. 12, pp. 1604–1613, 2014.

[69] L. Shen, E. Shen, Y. Luo, X. Yang, X. Hu, X. Zhang, Z. Tai, and J. Wang, "Towards natural language interfaces for data visualization: A survey," *IEEE Transactions on Visualization and Computer Graphics*, vol. 29, no. 6, pp. 3121–3144, 2022.

[70] G. Michelet and F. Breitinger, "Chatgpt, llama, can you write my report? an experiment on assisted digital forensics reports written using (local) large language models," *Forensic Science International: Digital Investigation*, vol. 48, p. 301683, 2024.

[71] J. W. Streefkerk, C. Wiering, M. van Esch-Bussemakers, and M. Neerincx, "Effects of presentation modality on team awareness and choice accuracy in a simulated police team task," in *Proceedings of the Human Factors and Ergonomics Society Annual Meeting*, vol. 52, no. 4. SAGE Publications Sage CA: Los Angeles, CA, 2008, pp. 378–382.

[72] E. Bertini and D. Lalanne, "Surveying the complementary role of automatic data analysis and visualization in knowledge discovery," in *Proceedings of the ACM SIGKDD Workshop on Visual Analytics and Knowledge Discovery: Integrating Automated Analysis with Interactive Exploration*, 2009, pp. 12–20.

[73] J. Heer and B. Shneiderman, "Interactive dynamics for visual analysis: A taxonomy of tools that support the fluent and flexible use of visualizations," *Queue*, vol. 10, no. 2, pp. 30–55, 2012.

[74] P. Isenberg, N. Elmqvist, J. Scholtz, D. Cernea, K.-L. Ma, and H. Hagen, "Collaborative visualization: Definition, challenges, and research agenda," *Information Visualization*, vol. 10, no. 4, pp. 310–326, 2011.

[75] A. Rind, T. D. Wang, W. Aigner, S. Miksch, K. Wongsuphasawat, C. Plaisant, B. Shneiderman *et al.*, "Interactive information visualization to explore and query electronic health records," *Foundations and Trends® in Human–Computer Interaction*, vol. 5, no. 3, pp. 207–298, 2013.

[76] G. Lee, R. Gommers, F. Waselewski, K. Wohlfahrt, and A. O'Leary, "Pywavelets: A python package for wavelet analysis," *Journal of Open Source Software*, vol. 4, no. 36, p. 1237, 2019.

[77] A. Lundgard, C. Lee, and A. Satyanarayan, "Sociotechnical considerations for accessible visualization design," in *2019 IEEE Visualization Conference (VIS)*. IEEE, 2019, pp. 16–20.

[78] S. Few, *Now you see it: simple visualization techniques for quantitative analysis*. Analytics Press, 2009.

[79] E. P. Baumer and M. S. Silberman, "When the implication is not to design (technology)," in *Proceedings of the SIGCHI Conference on Human Factors in Computing Systems*, 2011, pp. 2271–2274.

[80] [Authors], "Visual large language models for generalized and specialized applications," *arXiv preprint*, 2025.

[81] F. Sebastiani, "Classification of text, automatic," *The encyclopedia of language and linguistics*, vol. 14, pp. 457–462, 2006.

[82] B. Liu, *Sentiment analysis: Mining opinions, sentiments, and emotions*. Cambridge university press, 2020.

[83] M. Ehrmann, A. Hamdi, E. L. Pontes, M. Romanello, and A. Doucet, "Named entity recognition and classification in historical documents: A survey," *ACM Computing Surveys*, vol. 56, no. 2, pp. 1–47, 2023.

[84] A. Mendhakar and D. HS, "Parts-of-speech (pos) analysis and classification of various text genres," *Corpus-based Studies across Humanities*, vol. 1, no. 1, pp. 99–131, 2024.

[85] S. Sree Harsha, K. Krishna Swaroop, and B. Chandavarkar, "Natural language inference: Detecting contradiction and entailment in multilingual text," in *International Conference on Information Processing*. Springer, 2021, pp. 314–327.

[86] P. Gupta and V. Gupta, "A survey of text question answering techniques," *International Journal of Computer Applications*, vol. 53, no. 4, 2012.

[87] L. Xiao and X. Chen, "Enhancing llm with evolutionary fine tuning for news summary generation," *arXiv preprint arXiv:2307.02839*, 2023.

[88] A. Hendy, M. Abdelrehim, A. Sharaf, V. Raunak, M. Gabr, H. Matsushita, Y. J. Kim, M. Afify, and H. H. Awadalla, "How good are gpt models at machine translation? a comprehensive evaluation," *arXiv preprint arXiv:2302.09210*, 2023.

[89] G. Z. Higginbotham and N. S. Matthews, "Prompting and in-context learning: Optimizing prompts for mistral large," 2024.

[90] I. O. William and M. Altamimi, "Large language model for creative writing and article generation," *International Journal of Advanced Natural Sciences and Engineering Researches*, pp. 741–748, 2024.

[91] S. Venkatraman, N. I. Tripto, and D. Lee, "Collabstory: Multi-llm collaborative story generation and authorship analysis," *arXiv preprint arXiv:2406.12665*, 2024.

[92] J. Li, N. Huo, Y. Gao, J. Shi, Y. Zhao, G. Qu, B. Qin, X. Li, C. Ma, J.-G. Lou *et al.*, "Benchmarking intelligent llm agents for conversational data analysis."

[93] J. Gorniak, Y. Kim, D. Wei, and N. W. Kim, "Vizability: Enhancing chart accessibility with llm-based conversational interaction," in *Proceedings of the 37th Annual ACM Symposium on User Interface Software and Technology*, 2024, pp. 1–19.

[94] K. B. Buldu, S. Özdel, K. H. C. Lau, M. Wang, D. Saad, S. Schönborn, A. Boch, E. Kasneci, and E. Bozkir, "Cuify the xr: An open-source package to embed llm-powered conversational agents in xr," in *2025 IEEE International Conference on Artificial Intelligence and eXtended and Virtual Reality (AIxVR)*. IEEE, 2025, pp. 192–197.

[95] S. Brade, B. Wang, M. Sousa, S. Oore, and T. Grossman, "Promptify: Text-to-image generation through interactive prompt exploration with large language models," in *Proceedings of the 36th Annual ACM Symposium on User Interface Software and Technology*, 2023, pp. 1–14.

[96] A. McCallum, "Information extraction: Distilling structured data from unstructured text," *Queue*, vol. 3, no. 9, pp. 48–57, 2005.

[97] L. Wang, S. Zhang, Y. Wang, E.-P. Lim, and Y. Wang, "Llm4vis: Explainable visualization recommendation using chatgpt," *arXiv preprint arXiv:2310.07652*, 2023.

[98] V. Dibia, "Lida: A tool for automatic generation of grammar-agnostic visualizations and infographics using large language models," *arXiv preprint arXiv:2303.02927*, 2023.

[99] P. Maddigan and T. Susnjak, "Chat2vis: Generating data visualisations via natural language using chatgpt, codex and gpt-3 large language models," *Ieee Access*, 2023.

[100] Julius AI, "Julius AI: Advanced Data Analytics Platform," https://julius.ai/, accessed: 2025-03-15.

[101] Zoho Analytics, "Zoho Analytics: Self-Service BI and Data Analytics Platform," https://www.zoho.com/analytics/, accessed: 2025-03-15.

[102] J. Yang, P. F. Gyarmati, Z. Zeng, and D. Moritz, "Draco 2: An extensible platform to model visualization design," in *2023 IEEE Visualization and Visual Analytics (VIS)*. IEEE, 2023, pp. 166–170.
[103] A. Tushar, "MatplotLLM," https://emacsconf.org/2023/talks/matplotllm/, 2023, accessed: Mar. 13, 2025.
[104] T. Salesforce, "Tableau shows what's next for data analytics and ai at dreamforce 2023," https://www.tableau.com/blog/tableau-data-analytics-and-ai-dreamforce-2023, 2023.
[105] Y. Zhao, J. Wang, L. Xiang, X. Zhang, Z. Guo, C. Turkay, Y. Zhang, and S. Chen, "Lightva: Lightweight visual analytics with llm agent-based task planning and execution," *IEEE Transactions on Visualization and Computer Graphics*, 2024.
[106] A. Liew and K. Mueller, "Using large language models to generate engaging captions for data visualizations," *arXiv preprint arXiv:2212.14047*, 2022.
[107] H.-K. Ko, H. Jeon, G. Park, D. H. Kim, N. W. Kim, J. Kim, and J. Seo, "Natural language dataset generation framework for visualizations powered by large language models," in *Proceedings of the 2024 CHI Conference on Human Factors in Computing Systems*, 2024, pp. 1–22.
[108] K. Choe, C. Lee, S. Lee, J. Song, A. Cho, N. W. Kim, and J. Seo, "Enhancing data literacy on-demand: Llms as guides for novices in chart interpretation," *IEEE Transactions on Visualization and Computer Graphics*, 2024.
[109] Y. Han, C. Zhang, X. Chen, X. Yang, Z. Wang, G. Yu, B. Fu, and H. Zhang, "Chartllama: A multimodal llm for chart understanding and generation," *arXiv preprint arXiv:2311.16483*, 2023.
[110] Z. Wang, M. Xia, L. He, H. Chen, Y. Liu, R. Zhu, K. Liang, X. Wu, H. Liu, S. Malladi *et al.*, "Charxiv: Charting gaps in realistic chart understanding in multimodal llms," *Advances in Neural Information Processing Systems*, vol. 37, pp. 113 569–113 697, 2024.
[111] K. Gu, R. Shang, T. Althoff, C. Wang, and S. M. Drucker, "How do analysts understand and verify ai-assisted data analyses?" *arXiv preprint arXiv:2309.10947*, 2023.
[112] S. Zhang, L. Wang, T. J.-J. Li, Q. Shen, Y. Cao, and Y. Wang, "Chartifytext: Automated chart generation from data-involved texts via llm," *arXiv preprint arXiv:2410.14331*, 2024.
[113] M. M. Rashid, H. K. Jahan, A. Huzzat, R. A. Rahul, T. B. Zakir, F. Meem, M. S. H. Mukta, and S. Shatabda, "Text2chart: A multi-staged chart generator from natural language text," in *Pacific-Asia Conference on Knowledge Discovery and Data Mining*. Springer, 2022, pp. 3–16.
[114] Y. Tian, W. Cui, D. Deng, X. Yi, Y. Yang, H. Zhang, and Y. Wu, "Chartgpt: Leveraging llms to generate charts from abstract natural language," *IEEE Transactions on Visualization and Computer Graphics*, 2024.
[115] Y. Jin, A. Carrasco-Revilla, and M. Chen, "igaiva: Integrated generative ai and visual analytics in a machine learning workflow for text classification," *arXiv preprint arXiv:2409.15848*, 2024.
[116] [Authors], "Leva: Using large language models to enhance visual analytics," *arXiv preprint*, 2024.
[117] A. Masry, D. X. Long, J. Q. Tan, S. Joty, and E. Hoque, "Chartqa: A benchmark for question answering about charts with visual and logical reasoning," *arXiv preprint arXiv:2203.10244*, 2022.
[118] S. E. Kahou, V. Michalski, A. Atkinson, Á. Kádár, A. Trischler, and Y. Bengio, "Figureqa: An annotated figure dataset for visual reasoning," *arXiv preprint arXiv:1710.07300*, 2017.
[119] Y. Luo, N. Tang, G. Li, C. Chai, W. Li, and X. Qin, "Synthesizing natural language to visualization (nl2vis) benchmarks from nl2sql benchmarks," in *Proceedings of the 2021 International Conference on Management of Data*, 2021, pp. 1235–1247.
[120] H. Li, Y. Wang, and H. Qu, "Where are we so far? understanding data storytelling tools from the perspective of human-ai collaboration," *arXiv preprint arXiv:2309.15723*, 2023.
[121] M.-H. Hong and A. Crisan, "Conversational ai threads for visualizing multidimensional datasets," *arXiv preprint arXiv:2311.05590*, 2023.
[122] L. Christino and F. V. Paulovich, "Chatkg: Visualizing temporal patterns as knowledge graph," *The Eurographics Association*, 2023.
[123] F. Trebuňa and O. Dušek, "Visuallm: Easy web-based visualization for neural language generation," in *Proceedings of the 16th International Natural Language Generation Conference: System Demonstrations*, 2023, pp. 6–8.
[124] Z. Li, H. Miao, V. Pascucci, and S. Liu, "Visualization literacy of multimodal large language models: A comparative study," 2024. [Online]. Available: https://arxiv.org/abs/2407.10996
[125] L. Schulze Buschoff, E. Akata, M. Bethge *et al.*, "Visual cognition in multimodal large language models," *Nature Machine Intelligence*, vol. 7, pp. 96–106, 2025.
[126] A. Mishra, U. Soni, A. Arunkumar, J. Huang, B. C. Kwon, and C. Bryan, "Promptaid: Prompt exploration, perturbation, testing and iteration using visual analytics for large language models," *arXiv preprint arXiv:2304.01964*, 2023.

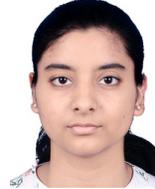

**Navya Sonal Agarwal** Navya Sonal Agarwal is a final-year undergraduate student majoring in Computer Science and Engineering at Kalinga institute of Industrial Technology Bhubaneswar, Odisha, India. She is currently pursuing research in the field of artificial intelligence with a particular interest in large language models and visual analytics. Her research primarily explores interdisciplinary approaches combining machine learning, visualization, and natural language processing.

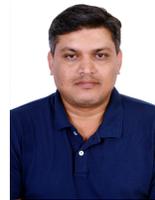

**Sanjay Kumar Sonbhadra** Sanjay Kumar Sonbhadra is presently working as an Assistant Professor in the Computer Science and Engineering Department of ITER, Shiksha 'O' Anusandhan, Bhubaneswar, Odisha, India. He is mainly working on One-class classification, Anomaly detection, Dimensionality reduction and Training sample selection techniques to handle large-scale data processing and visualization.